\documentclass[traditabstract]{aa}

\usepackage{graphicx}
\usepackage{epsfig}
\usepackage{amsmath,amssymb,amsfonts}

\usepackage[latin1]{inputenc}
\usepackage[english]{babel}

\newcommand{\etal}{et al. }
\newcommand{\dd}{\mathrm{d}}

\newcommand{\mat}{\mathrm{m}}

\newcommand{\Om}{\Omega_\mat}

\newcommand{\be} {\begin{equation}}
\newcommand{\ee} {\end{equation}}
\newcommand{\Qa}{Q_\mathrm{a}}
\newcommand{\etaa}{\eta_\mathrm{a}}
\newcommand{\Sa}{\Sigma_\mathrm{a}}
\newcommand{\Ga}{\Gamma_\mathrm{a}}

\begin{document}

 \title{COSMOS\thanks{Based on observations made with the NASA/ESA {\em Hubble Space Telescope}, obtained from the data archives at the Space Telescope European Coordination Facility and the Space Telescope Science Institute, which is operated by the Association of Universities for Research in Astronomy, Inc., under NASA contract NAS 5-26555.} weak-lensing
   constraints on modified gravity}


 \author{ 
   Ismael Tereno\inst{1}
   \and 
   Elisabetta Semboloni\inst{2}
   \and
   Tim Schrabback\inst{2}
   }

 \offprints{{\tt tereno@fc.ul.pt}}

 \institute{
 Centro de Astronomia e Astrof\'{\i}sica da Universidade de Lisboa, Tapada da Ajuda, 1349-018 Lisboa, Portugal \and
 Leiden Observatory, Leiden University, Niels Bohrweg 2, NL-2333 CA Leiden, The Netherlands 
}

   \date{Received ; accepted}



\abstract{

{The observed acceleration of the universe, explained through dark energy, could alternatively be explained through
a modification of gravity that would also induce modifications in the
evolution of cosmological perturbations.}

{We use new weak lensing data from the COSMOS survey to test for deviations from General Relativity.
The departure from GR is parametrized in a model-independent way that
consistently parametrizes the two-point cosmic shear amplitude and growth.} 

{Using CMB priors, we perform a likelihood analysis. We find constraints on the amplitude of the signal that do not indicate a
deviation from General Relativity.}

}


\maketitle


\section{Introduction}\label{sec1}

The $\Lambda$CDM model is currently the best fit to the available cosmological data. In this framework, the period of accelerated expansion of the Universe is due to dark energy, a dominant component of the energy density with a negative equation of state. A possible alternative, worth to investigate observationally, is to replace the standard gravitational theory of General Relativity (GR) by a modification that admits self-accelerating solutions. Such gravity theory must converge to GR both at high redshift, when the expansion is not accelerated, and on small scales, since GR has consistently passed local tests of gravity (\cite{will}). 

The various classes of modified gravity and dark energy models (\cite{uzan}) cannot be distinguished based only on observations of the cosmological background. They are only distinguishable due to their different predictions for cosmological structure formation (\cite{zhang}). 
The evolution of cosmological background and perturbations have been studied in various classes of physically motivated modified gravity theories. Most prominent today are theories where the modifications arise from extra dimensions, with matter fields confined to a 4-dimensional brane embedded in a higher dimensional bulk, such as the Dvali-Gabadadze-Porrati (DGP) model (\cite{dvali}; \cite{deffayet}; \cite{lue}), and also the so-called f(R) theories (\cite{carroll}; \cite{songfr}). In the latter, the field equations are modified because the corresponding Einstein action no longer depends linearly on the Ricci scalar (R) but there are extra terms on R. 

The standard way of studying cosmological structure formation is to introduce scalar perturbations in the Friedmann-Robertson-Walker (FRW) spacetime metric. In the Newtonian or longitudinal gauge (\cite{mukhanov}), where by construction the metric remains diagonal, there are only two scalar perturbations: the scalar fields $\Psi$ (the Newtonian potential) and $\Phi$ (the spatial curvature potential). They are sourced by the cosmological fluid perturbations contained in the stress-energy tensor. Conservation of the stress-energy tensor imposes evolution equations for its components. In addition, the equations of the theory of gravitation, relating the metric components with the energy components, bring extra evolution and constraint equations.  

In modified gravity, structure formation has been studied with various parametric and non-parametric approaches (\cite{bertschinger}; \cite{silvestri}). Parametric approaches include direct parametrization of the linear growth (\cite{linder}; \cite{lindercahn}); the so-called parametrized post-Friedmann framework (PPF) (\cite{husawicki}) and several related phenomenological post-GR parametrizations, which give a model-independent description of deviations to GR in a way that is convenient for the testing of cosmological predictions (\cite{caldwell}; \cite{amendola}; \cite{daniel08}; \cite{daniel10}; \cite{pogosian}); and phenomenological applications to parametrizations of particular theories of modified gravity (\cite{acquaviva}; \cite{zhaoa}; \cite{songbd}).

The various metric and energy perturbation variables in modified gravity scenarios may be measured by different probes (\cite{huterer}; \cite{jainz}; \cite{schmidt}; \cite{song}; \cite{guzik}; \cite{beynon}). In particular, weak lensing by large-scale structure (cosmic shear), which probes the sum $\Phi + \Psi$, is well suited for such studies (\cite{uzanb}; \cite{knox}). In cosmic shear measurements, the gravitational information is contained in the equations relating the measured matter density perturbations with the two potentials, or equivalently in one of those equations (a generalized Poisson equation) and an equation relating the two potentials (the anisotropy equation). A parametrization of those two relations is thus well suited for cosmic shear measurements, which will provide a degenerate estimation of those two quantities. This is the approach we follow in this work.

We compute cosmic shear two-point correlation models in tomographic bins, using standard cosmological parameters plus two modified gravity parameters, and analyse them against the new COSMOS tomographic cosmic shear measurements of Schrabback \etal (2010). Other cosmic shear data have already been used in recent studies of modified gravity, using similar or different parametrizations (\cite{dore}; \cite{thomas}; \cite{daniel09}; \cite{bean}; \cite{zhaotom}; \cite{daniellinder}). 
The standard parameters are constrained by using WMAP7-year CMB data (\cite{komatsu}). Our goal is to study the impact of this phenomenological model in the degeneracy between the amplitude of the cosmic shear signal and the linear growth of structure. In other words, to investigate if, given the strong cosmological priors imposed by the WMAP7 results, there is room to compensate a possible alteration of the predicted redshift-dependent amplitude of the cosmic shear effect with a modification of the growth of matter perturbations, which, phenomenologically, would keep open the possibility of an alternate gravity theory as an origin for an effective dark energy behaviour.

In the next section, we present the parametrization of modified gravity used. The new COSMOS dataset is described in Sect.~3. The statistical analysis and its results are presented and discussed in Sects.~4 and 5. We conclude in Sect.~6 with a reminder of the main assumptions made.

\section{Parametrization}\label{sec2}

Images of background galaxies are distorted (by convergence and shear) in the perturbed FRW spacetime. In the Newtonian gauge, considering scalar perturbations and a flat background, this spacetime is described by
\be
\dd s^2=-\left(1+2\Psi(x,t)\right)\dd t^2+a^2(t)\left(1-2\Phi(x,t)\right)\delta_{ij}\dd x^i \dd x^j\,.
\ee
The local deflection of null geodesics, with respect to unperturbed ones, depends on the traveling time of a light ray, and thus on the sum of the metric perturbations,
\be
\frac{\dd t}{\dd x}\approx\,1-(\Psi+\Phi)\,.
\ee
Integrating the multiple deflections over cosmological distances, and introducing the lens equation, the cosmic convergence is written, to first order in the amplification matrix and for a flat Universe, as (\cite{bartelmann})
\be
\kappa(\theta,w)=\frac{1}{2}\int_0^{w}\dd w'\,\frac{w-w'}{w}w'
\nabla^2\left[\Psi+\Phi\right](w'\theta,w')\,,
\label{eq:kappa}
\ee
where $(\theta,w)$ are comoving spatial coordinates and the Laplacian of the lenses potentials at comoving distances $w'$ is integrated up to the source at distance $w$. 

The equations of the theory of gravitation at perturbation level relate the metric perturbations (the potentials $\Psi$, $\Phi$) to the perturbations of the cosmological fluid (matter density contrast $\delta$, pressure perturbation $\delta p$, fluid velocity $v$, and stress or anisotropic pressure $\sigma$). In GR these equations are the linearized Einstein equations. The phenomenological approach parametrizes deviations from GR by introducing extra functions in the linearized Einstein equations. 

Following Ma \& Bertschinger (1995), a combination of the $0-0$ and $0-i$ linearized Einstein equations relates $\Phi$ with the comoving matter density perturbation: it is the generalized Poisson equation. In Fourier space, with $k^2$ being the spatial Laplacian, it writes,
\be
k^2\Phi=-4\pi\,G\,a^2\rho\Delta=-\frac{3}{2}\frac{\Om H_0^2\,\Delta}{a}\,,
\label{eq:poisson}
\ee
where $\Delta$ is the comoving density perturbation $\Delta=\delta+3H(1+w)v/k^2$, $w$ is the equation-of-state parameter $w=p/\rho$, and the second equality is written for dark matter inserting $\Om$. The deviation from GR is parametrized by introducing the {\em mass screening} phenomenological post-GR function $Q(a,k)$, replacing the gravitational constant by an effective function,
\be
G_{\rm eff}=G\,Q\,\Rightarrow\,k^2\Phi=-4\pi\,G\,Q\,a^2\rho\Delta\,.
\label{eq:poisson2}
\ee
Notice that a hypothetical clustering dark energy component would also add a contribution to the Poisson equation. Factorizing out the dark matter density contrast, a multiplicative term, function of $\delta_{\rm DE}/\delta$, would appear, and play the same role as $Q$.

The $i-j$ Einstein equation gives a second constraint, relating the difference $\Phi-\Psi$ with the stress $\sigma$: it is the anisotropy equation,
\be
k^2(\Phi-\Psi)=12\pi\,G\,a^2\rho\,(1+w)\sigma.
\label{eq:gslip}
\ee
In GR the two potentials are identical.
The deviation from GR is parametrized by introducing the {\em gravitational slip} phenomenological post-GR function $\eta(a,k)$,
\be
\Psi=(1+\eta)\Phi\,\Rightarrow\,k^2(\Phi-\Psi)=4\pi\,G\,Q\eta\,a^2\rho\Delta\,.
\label{eq:gslip2}
\ee
Again, a hypothetical dark energy component with anisotropic stress $\sigma$ would have the same effect as $Q\eta$.

The two remaining independent Einstein equations, for example $0-i$ and $i-i$, are evolution equations for $\Psi$ and $\Phi$, and will not be needed when considering the quasi-static approximation, valid for sub-horizon scales.

The linearized Einstein equations allow thus for 2 independent post-GR functions.
From Eqs.~(\ref{eq:poisson2}) and (\ref{eq:gslip2}), we can also write modified equations for the Laplacians of $\Phi+\Psi$ and $\Psi$.
They are
\be
k^2(\Phi+\Psi)=-8\pi\,G\,Q\left(1+\frac{\eta}{2}\right)\,a^2\rho\Delta\
\label{eq:modgrav3}
\ee
and
\be
k^2\Psi=-4\pi\,G\,Q(1+\eta)\,a^2\rho\Delta\,.
\label{eq:modgrav4}
\ee
These equations introduce two derived post-GR functions that are useful for cosmic shear studies: $\Sigma(a)=Q(a)(1+\eta(a)/2)$ and $\Gamma(a)=Q(a)(1+\eta(a))$.

One approach, taken e.g by Daniel \& Linder (2010) and Zhao \etal (2010), to study the post-GR modifications as function of scale and redshift is to parametrize them with various constant parameters in redshift and scale bins. Here, we consider them to be scale-independent in the limited range of scales provided by the COSMOS data, although it is clear that in a broad range of scales there is scale dependence at least to differentiate the GR behaviour on solar system scales from the possible modifications on cosmological scales. 

Regarding the redshift dependence, we adopt a functional form using the scaling argument of Caldwell \etal (2007), according to which the departure from GR originates an effective smooth dark energy $\Lambda$ with constant $w=-1$. The deviations in the growth history scale with the deviations in the expansion history, such that they do not turn on too early, deviating from GR at early times, or too late. In this approach, the background evolution is identical to $\Lambda$CDM, no further modification is assumed in the Friedmann equations, and the post-GR functions take the GR values at $a\ll 1$ ($Q=1$, $\eta=0$ or $\Sigma=1$, $\Gamma=1$) and deviate from GR proportionally to $\rho_{\rm DE}(a)/\rho_{\rm m}(a)$. This implies that
\be
Q(a)=1+\Qa\,a^3\,, \;\; \eta(a)=\etaa\,a^3\,.
\ee
This dependence for the fundamental functions $Q(a)$ and $\eta(a)$ implies a different redshift dependence for the derived functions $\Sigma(a)$ and $\Gamma(a)$. Only in the case of small parameter values, $(\Qa<1$, $\etaa<1)$, do the derived functions have a similar scaling,
\be
\Sigma(a)\approx 1+\Sa\,a^3\,, \;\; \Gamma(a)\approx \Ga\,a^3\,,
\ee
where $\Sa=\Qa+\etaa/2$ and $\Ga=\Qa+\etaa$.
We will work in this ansatz, which reduces the range of values allowed for the post-GR parameters. In particular, we will consider models such that the difference between the two potentials is less than $50\%$ and the effective gravitational constant does not differ more than $50\%$ from the GR value. Large deviations are better studied with direct parametrizations in redshift bins, in the absence of a well-motivated functional form. Also, the approach of modifying the linearized Einstein equations should break down for too large deviations.

Going back to the convergence, Eq.~(\ref{eq:kappa}), we compute its power spectrum (\cite{bartelmann}) and, inserting Eqs.~(\ref{eq:poisson2}) and (\ref{eq:gslip2}), obtain the cosmic shear power spectrum in the following form,
\be
P_\gamma^{ij}(\ell)=\frac{9}{4}\,\Om^2 H_0^4\int_0^{w_{\rm{h}}}dw\frac{g_i(w)g_j(w)}{a^2(w)}\,\Sigma^2(a) P_\delta\left(\frac{\ell}{w}\,,w\right).
\label{eq:pkappa}
\ee
The amplitude of the power spectrum depends on the derived function $\Sigma(a)$.
This equation considers the correlation of source galaxies from two redshift bins, $i,j$, with redshift distributions $p\,(w)$, which are lensed with geometrical efficiency factors given by,
\be
g_i(w)=\int_w^{w_h}\dd w' p_i(w')\frac{w'-w}{w'}\,.
\ee 

Deviations to GR thus modify the cosmic shear amplitude via $\Sigma$. They further impact the cosmic shear power spectrum via a modification of the matter power spectrum, $P_\delta$, present in Eq.~(\ref{eq:pkappa}). The conservation of stress-energy provides the evolution equations for the matter perturbations. The evolution equations for $\delta$ and $v$ depend on the stress-energy components, the time evolution of the potentials and on the Laplacian of the Newtonian potential (\cite{ma}).
Inserting the constraint equations Eqs.~(\ref{eq:poisson2}) and (\ref{eq:gslip2}), and writing for the comoving density $\Delta$, a great number of terms arise. The solution is a scale-dependent growth in the linear regime, even for scale-independent post-GR parameters (\cite{daniel10}). 

For the case of sub-horizon scales, assuming time derivatives of the perturbations are negligible compared to the spatial derivatives, and $\Delta\approx \delta$, the evolution equation of the cold dark matter perturbation greatly simplifies. It only contains a Hubble drag term and a term on $k^2\Psi$ which means, from Eq.~(\ref{eq:modgrav4}), that the growth depends directly on the function $\Gamma(a)$. 

In this regime, the growth solution can be cast in a form that explicitly separates the expansion history and gravitational dependencies of the growth (\cite{linder}), 
\be
\frac{\dd \ln(\delta/a)}{\dd \ln(a)}=\Om(a)^\gamma-1\,.
\label{eq:growth}
\ee
This introduces the growth parameter, which contains the gravitational dependence, and has the value $\gamma=6/11$ in GR (\cite{wang}). In the case of $\Lambda$CDM background evolution and scaling evolution of $\Gamma(a)$, it takes the value (\cite{lindercahn}; \cite{daniel10}),
\be
\gamma=\frac{6}{11}\left[1-\frac{\Gamma_{\rm a}}{2}\frac{\Om}{1-\Om}\right]\,.
\label{eq:gamma}
\ee

\section{COSMOS cosmic shear data}\label{sec3}

The new COSMOS cosmic shear catalogue (\cite{schrabback}) (hereafter S10) contains around $450\,000$ galaxies with magnitude $i<26.7$ in an area of 1.64$\,\deg^2$, corresponding to a density of 76 galaxies$/{\rm arcmin}^2$. 

The cosmic shear analysis of S10 used the COSMOS-30 photometric redshift catalogue (\cite{ilbert}), which provides individual photometric redshifts for galaxies of magnitude $i<25$. Those were grouped in 5 redshift bins. The highest magnitude galaxies $(i>24)$  of the lowest redshift bin $(z<0.6)$ suffer from catastrophic errors due to the dificulty to locate the 4000 \AA\ break in their spectra. They are thus likely to contain high-redshift galaxies and were removed from the analysis. In addition, the redshift distribution of galaxies with $i>25$, for which individual redshifts were not available, was estimated in S10 and those galaxies were placed in a single wide redshift bin. Table~\ref{tab:zbins} shows the 6 redshift bins used, containing the remaining roughly $420\,000$ galaxies.

\begin{table}[tb]
\begin{center}
\caption{Description of the redshift bins used in the analysis, with their amount of contributing galaxies and mean redshifts. From Schrabback \etal (2010).
\label{tab:zbins}
}
\vspace{0.2cm}
\begin{tabular} {cccccc}
\hline
Bin & $z_\mathrm{min}$ & $z_\mathrm{max}$ & $N$&  $\langle z \rangle$ \\
\hline
1 & 0.0 & 0.6 & 22\,294 &  0.37\\
2 & 0.6 & 1.0 & 58\,194 &  0.80\\
3 & 1.0 & 1.3 & 36\,382 &  1.16\\
4 & 1.3 & 2.0 & 25\,928 &  1.60\\
5 & 2.0 & 4.0 & 21\,718 &  2.61\\
\hline
6 & 0.0 & 5.0 &  251\,958  & $1.54\pm0.15$  \\
\hline
\end{tabular}
\end{center}
\end{table}

The data used in this work consists on a vector of shear correlations $\xi_+$ and $\xi_-$ computed with the estimator given in Eq.~(13) of S10. We use all the cross-correlations between redshift bins (12, 13, 14, 15, 23, 24, 25, 34, 35, 45, 16, 26, 36, 46, 56) and the auto-correlation of the wide sixth bin. The auto-correlations of the narrow bins (11, 22, 33, 44, 55) were not used to minimize the effect of intrinsic alignments. In all cases, the shear correlations were computed on 5 angular separation bins ranging from 0.7 to 21 arcmin. The data vector has thus $(15+1)\times 5 \times 2=160$ components. S10 found the B-modes to be consistent with zero at all scales. In this case, we work directly with the $\xi_\pm$ correlations and there is no need to make an E/B mode decomposition of the shear field, which has the disadvantage of being non-local. 

The $160 \times 160$ covariance matrix of the shear correlations was estimated in S10 from 288 realisations of  
COSMOS-like fields obtained from ray-tracing through the Millenium simulation (\cite{hilbert}). The inverse covariance matrix, needed for the likelihood analysis, was debiased by applying the correction of Hartlap \etal (2007).

\section{Analysis}\label{sec4}

\begin{figure}
\includegraphics[width=7cm]{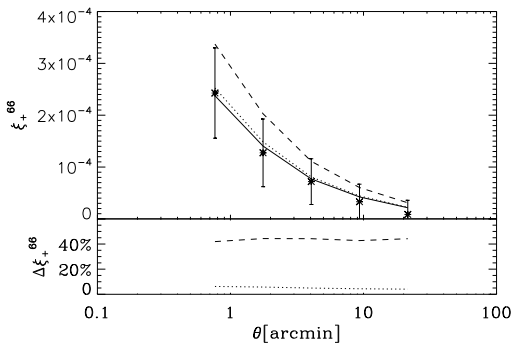} 
  \caption{Correlation function $\xi_+^{66}$. Top panel shows the measurement (points with error bars), the best-fit GR model for a non-tomographic analysis ($\Omega_m=0.27$, $\sigma_8=0.76$) (solid), modified gravity models with the same standard parameters but ($\Sigma=1.6$, $\gamma=0.55$) (dashed), or ($\Sigma=1$, $\gamma=0.65$) (dotted). Bottom panel shows the deviation between the same dashed and dotted models with respect to the GR model.}
  \label{fig:2pt}
\end{figure}

We build models on a three-dimensional grid consisting of $(\Qa,\etaa)$ and a redshift-calibration parameter $f_z$. 

The parameter $f_z$ is a nuisance parameter that allows to include an uncertainty on the redshift calibration of bin 6. Following S10, we marginalize over this uncertainty by including shifted redshift distributions on the grid of models, such that each model is evaluated assuming various redshift distributions $p\,(f_z\,z)$ in the range $0.9<f_z<1.1$.

We choose to work with the two original phenomenological parameters, to be able to have an uniform coverage of the $(\Qa,\etaa)$ plane.
We probe a relatively narrow range of values: $-0.5<\Qa<0.5$ and $-0.5<\etaa<0.5$. As discussed in Sect.~2, this choice allows the redshift-dependence of the post-GR derived parameters to scale with the background expansion. The grid mapped on the plane of the derived parameters, the ones that have a direct impact on the cosmic shear signal, i.e., the amplitude $\Sa$ and the growth $\Ga$ or $\gamma$, introduces a correlation between these 2 parameters. Its range is such that $\Sigma(a=1)$ can deviate from the GR value, $\Sigma(a=1)=1$, by up to $70\%$, while $\gamma$ extends to a maximum deviation of $40\%$ of the GR value of $\gamma=0.55$. (Hereafter we denote the values of the post-GR parameters at $a=1$ by $Q$, $\eta$, $\Sigma$ and $\Gamma$). 

The grid is computed for each point of a sample of standard cosmological models built by thinning-out the $\Lambda$CDM WMAP-7 year Monte Carlo Markov Chains available in the WMAP archive \footnote{http://lambda.gsfc.nasa.gov/}.  In this way, the analysis has strong priors on the standard cosmological parameters. The average $\sigma_8$ and $\Omega_m$ values of the models used are $\sigma_8=0.807$ and $\Omega_m=0.271$.

For each model, we evaluate the various shear power spectra using Eq.~(\ref{eq:pkappa}). The matter power spectrum is computed using the transfer function of Eisenstein \& Hu (1998) and the linear growth of Eq.~(\ref{eq:growth}) with $\gamma$ given in Eq.~(\ref{eq:gamma}).
Since the small scales of the COSMOS data are in the quasi-linear and non-linear regime, we need a non-linear prescription to correct the linear power spectrum. We compute the non-linear power spectrum using the formula of Smith \etal (2003), which was built from N-body simulations of $\Lambda$CDM. Finally, the shear correlations are computed from
\be
\xi^{ij}_{+/-}(\theta)=\frac {1}{2\pi}\int_0^\infty \dd \ell\,\ell  J_{0/4}(\ell\theta)P_\kappa^{ij}(\ell)\,,
\ee
where $J_n$ is the $n^{\rm th}$order Bessel function of the first kind. 

As an example, Fig.~\ref{fig:2pt} shows the $\xi_+$ auto-correlation function for the wide redshift bin. In the models shown, $\Sigma$ and $\gamma$ take the extreme values used in the grid, when the other parameter is kept fixed at GR. Consistently with the approximations described before, the effect is scale-independent. We notice the impact of $\gamma$ is very small on the COSMOS angular scales whereas the impact of $\Sigma$ clearly dominates. The effect of the growth parameter would be more important on larger scales, due to the $k^2$ factor in the Poisson equation (see also \cite{daniel10}).

\begin{figure}
\includegraphics[width=4.5cm]{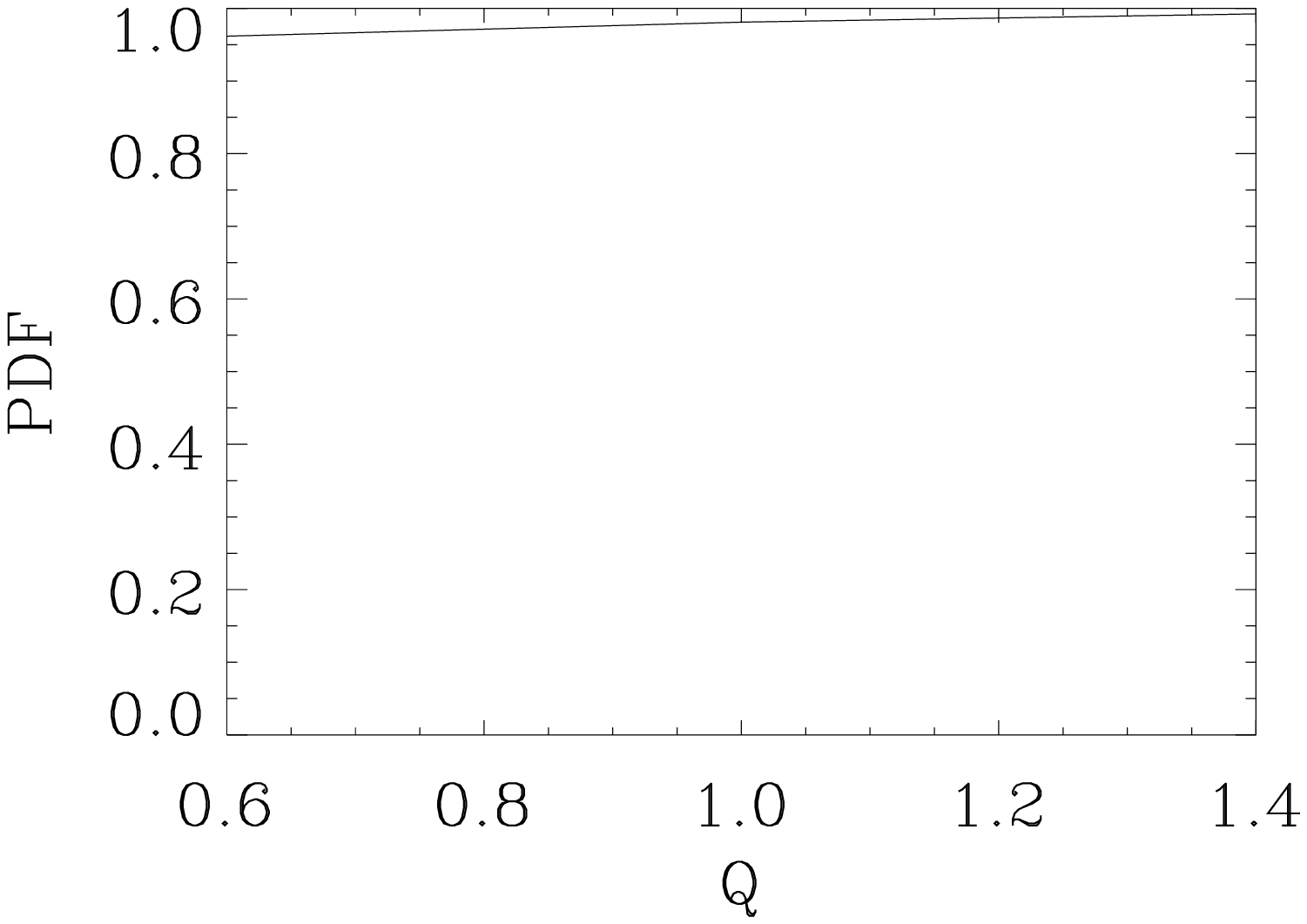} 
\hspace{-0.5cm}
\includegraphics[width=4.5cm]{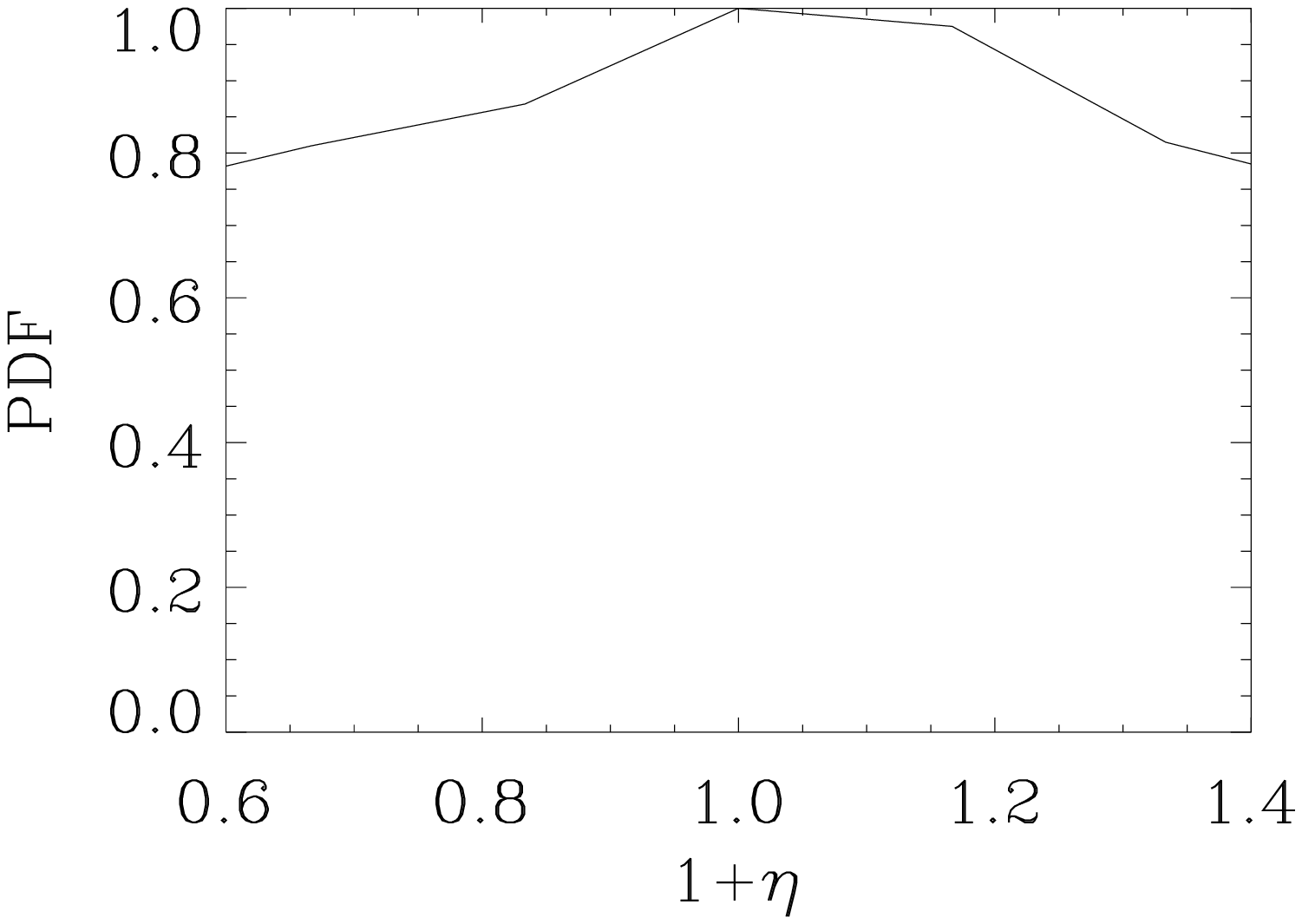} 
  \caption{Marginalised one-dimensional PDFs obtained for the phenomenological post-GR parameters.}
  \label{fig:1d}
\end{figure}

The likelihood of each model of the grid is evaluated with respect to the previously described COSMOS data vector of dimension 160. The likelihood values are then used to update the original weights of the chain, in order to obtain the joint CMB-cosmic shear weights according to the method of importance sampling (\cite{lewis}).

The resulting marginalised Probability Distribution Functions are shown in Fig.~\ref{fig:1d}, where it is clear that the
phenomenological parameters $Q$ and $\eta$ are not constrained by the data in the range of values probed. 

\section{Results and Discussion}\label{sec5}

\begin{figure}
\includegraphics[width=7cm]{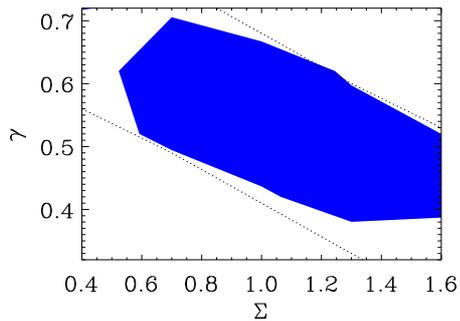} 
  \caption{The dotted lines show the limits of the probed region. The 68\% C.L. likelihood contour essentially fills up that region, showing that for these data the range of models chosen constitutes a significant prior.}
  \label{fig:2d2}
\end{figure}

We consider now the derived parameters, $\Sigma$ and $\gamma$, the ones that directly affect the cosmic shear signal. Fig.~\ref{fig:2d2} shows the result in the $(\Sigma, \gamma)$ plane. In this analysis, given the scale-independent approximations made (both in the choice of scale-independent parameters and in the use of small scales where the modifications to growth are scale-independent), the post-GR models differ only in their amplitudes and redshift-evolution of the amplitudes. $\Sigma$ and $\gamma$ are thus degenerate and we constrain the following combination of parameters:
\begin{equation}
\Sigma+2.3\,(\gamma-0.55)=0.99\pm 0.31\,(68\% {\rm C.L.}).
\end{equation}

This result combines information from the data, the cosmological prior, and the prior given by the range of the grid. The latter introduces a correlation between $\Sigma$ and $\gamma$ that arises from the fact that the uncorrelated grid on the $(Q,\eta)$ plane of the phenomenological parameters maps onto a correlated grid on the $(\Sigma,\gamma)$ plane, as shown in Fig.~\ref{fig:2d2}.  

\begin{figure}
\includegraphics[width=7cm]{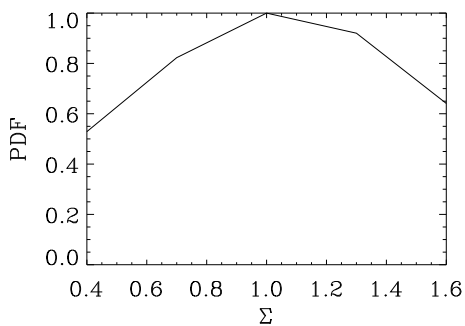} 
  \caption{Marginalised one-dimensional PDF of $\Sigma$.}
  \label{fig:1dsigma}
\end{figure}

The corresponding one-dimensional PDFs of both $\Sigma$ and $\gamma$ are centered on the GR values. This happens both for the marginalised cases and when fixing one of the parameters at the GR value. 
This shows that the amplitudes of the S10 COSMOS cosmic shear measurements agree with those of the WMAP 7-year CMB measurements. They both show a preference for the same value of $\sigma_8$. In other words, given the strong priors of WMAP the amplitude of the cosmic shear models is set, with the prefered ones lying close to GR. This can also be seen as an indirect test on the amplitude of systematics. Indeed, if there were systematics not accounted for in the data, or in the estimation of the redshift distribution, they would make the signal less compatible with the WMAP priors and could show up in the results as evidence for modified gravity. 
Fig.~\ref{fig:1dsigma} shows the PDF of $\Sigma$, marginalised over $\gamma$, which corresponds to a $1\sigma$ constraint of $\Sigma=1.02 \pm 0.38\,$.

\begin{figure}
\includegraphics[width=7cm]{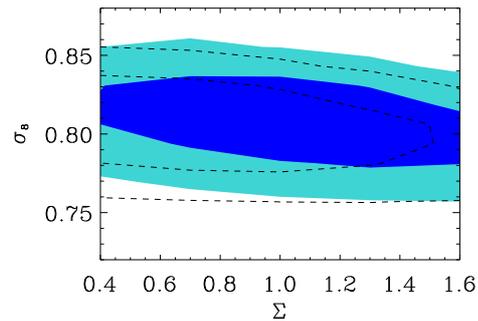} 
  \caption{Marginalised likelihood contours on the $(\Sigma,\sigma_8)$ plane at 68\% and 95\%
  confidence levels, obtained from the full analysis (filled contours), and using only the '66' correlation (open, dashed lines).}
  \label{fig:2d1}
\end{figure}

The finding that these post-GR parameters are centered on the GR values is quite robust, making use of the information on the redshift evolution of the amplitude from the tomographic correlations.
By repeating the analysis using only the wide bin auto-correlations (which contains $60\%$ of the galaxies) we found a slight change in the $(\sigma_8,\Sigma)$ degeneracy, shown in Fig.~\ref{fig:2d1}, and a shift on the mean of the posterior distribution of $\Sigma$ to $\bar\Sigma=0.87 \,$.

\section{Summary and Conclusions}\label{sec6}

We used the new COSMOS cosmic shear data to test possible deviations from General Relativity.
The departure from GR is parametrized in a phenomenological model-independent way extensively used in the literature.
The two phenomenological parameters allow to derive two other parameters that consistently parametrize the modifications to two-point cosmic shear amplitude and growth, and are scale-independent and have a scaling redshift dependence with the $\Lambda$CDM background. 

A more realistic parametrization should take into account three regimes of modified gravity (\cite{husawicki}), allowing for compatibility with the background expansion on large scales and agreement with GR on sub-galactic scales.  Hu \& Sawicki (2007) give a fitting formula for the non-linear matter power spectrum across the regimes. That formula was confirmed for particular classes of theories by perturbation theory (\cite{koyama}) and N-body simulations (\cite{oyaizu}). In particular for f(R), according to Koyama \etal (2009) it shows systematic deviations of at least 10\% from the formula of Smith \etal (2003), for scales smaller than $k\approx 0.8h\,{\rm M_{pc}}^{-1}$. COSMOS has an average source redshift of $z\approx 1.5$ and is mainly sensitive to lensing structures at around half comoving distance to the source. Therefore, this limiting scale corresponds to $\theta\approx 20\,{\rm arcmin}$ and the COSMOS range is affected. We do not apply this result to our model-independent study, but in general a systematic increase in the amplitudes of the models implies that cosmic shear becomes more sensitive to $\Sigma$, leading to a more precise determination of this parameter for the same data. 

Our results use WMAP priors obtained for $\Lambda$CDM models. This is consistent with using a $\Lambda$CDM background evolution. The CMB power spectrum is affected by the Integrated Sachs-Wolfe effect essentially at the recent epoch when the modified gravity effects become important and the potentials change significantly over time. This affects only the scales large enough compared with the photons crossing time. Since the effect is at low redshift those large physical scales correspond to very large angular scales. In this way, the peak structure is not changed and the parameters of modified gravity are not degenerate with the standard parameters at perturbation level, allowing to use $\Lambda$CDM priors.

We did not find evidence for inconsistency with GR, in agreement with other cosmological tests in the linear regime (see \cite{jain} for a review). In particular, we do not find the evidence for deviation reported by Daniel \& Linder (2010) when using small-scale CFHTLS data. Our result using the new tomographic correlations of COSMOS corroborates the results obtained by Daniel \& Linder (2010) using a different COSMOS catalogue (\cite{massey07}). We cannot however make a direct comparison of the constraints, since we assume a different redshift-dependence and different priors.

Our analysis relied on the tomographic amplitude of 2-pt cosmic shear, and did not allow to break the $\Sigma-\gamma$ degeneracy. 
The scale-dependence of the modifications, which is more distinctive on scales larger than those probed by COSMOS, should be explored with future wide high-precision weak lensing surveys. Such an analysis will have a better chance of reaching our original goal, i.e., to investigate if given strong cosmological priors there would be room to compensate a modification of the predicted amplitude of the cosmic shear signal with a modification of the growth of matter perturbations.

\begin{acknowledgements}
It is a pleasure to thank the COSMOS team for producing the data used in this work, in particular Jan Hartlap, Benjamin Joachimi, Martin Kilbinger and Patrick Simon. We thank Tommaso Giannantonio for fruitful discussions. IT is grateful to AIfA - University of Bonn and the Leiden Observatory for hospitality. IT is funded by an FCT (Portugal) fellowship and acknowledges support of the European Programmes ``DUEL'' RTN and ``CALICE'' ERG. TS acknowledges support from
the Netherlands Organization for Scientific Research (NWO).
\end{acknowledgements}



\begin{thebibliography}{}

\bibitem[Acquaviva \etal 2008]{acquaviva}
Acquaviva, V., Hajian, A., Spergel, D.~N \& Das, S. 2008, \prd \,78, 043514

\bibitem[Amendola \etal 2008]{amendola}
Amendola, L., Kunz, M. \& Sapone, D. 2008, \jcap \,4, 13

\bibitem[Bartelmann \& Schneider 2001]{bartelmann}
Bartelmann, M. \& Schneider, P. 2001, Phys. Rep. 340, 291

\bibitem[Bean \& Tangmatitham 2010]{bean}
Bean, R. \& Tangmatitham, M. 2010, \prd \,81, 083534

\bibitem[Beynon \etal 2010]{beynon}
Beynon, E., Bacon, D.~J \& Koyama, K. 2010, \mnras \,403, 353

\bibitem[Bertschinger \& Zukin 2008]{bertschinger}
Bertschinger, E. \& Zukin, P. 2008, \prd \,78, 024015

\bibitem[Caldwell \etal 2007]{caldwell}
Caldwell, R., Cooray, A. \& Melchiorri, A. 2007, \prd \,76, 023507

\bibitem[Carroll \etal 2004]{carroll}
Carroll, S., Duvvuri, V., Trodden, M. \& Turner, M.~S. 2004, \prd \,70, 043528

\bibitem[Daniel \& Linder 2010]{daniellinder}
Daniel, S.~F. \& Linder, E.~V. 2010, arXiv:1008.0397

\bibitem[Daniel \etal 2008]{daniel08}
Daniel, S.~F., Caldwell, R.~R., Cooray, A. \& Melchiorri, A. 2008, \prd \,77 103513

\bibitem[Daniel \etal 2009]{daniel09}
Daniel, S.~F., Caldwell, R.~R., Cooray, A., Serra, P. \& Melchiorri, A. 2009, \prd \,80, 023532

\bibitem[Daniel \etal 2010]{daniel10}
Daniel, S.~F., Linder, E.~V., Smith, T.~L. \etal 2010, \prd \,81, 123508

\bibitem[Deffayet \etal 2002]{deffayet}
Deffayet, C., Dvali, G. \& Gabadadze, G. 2002, \prd \,65, 044023

\bibitem[Dor{\'e} \etal 2007]{dore}
Dor{\'e}, O., Martig, M., Mellier, Y. \etal 2007, arXiv:0712.1599

\bibitem[Dvali \etal 2000]{dvali}
Dvali, G., Gabadadze, G. \& Porrati, M. 2000, Phys. Lett. B 485, 208

\bibitem[Eisentein \& Hu 1998]{eisenstein}
Eisenstein, D.~J. \& Hu, W. 1998, \apj \,496, 605 

\bibitem[Guzik \etal 2010]{guzik}
Guzik, J., Jain, B. \& Takada, M. 2010, \prd \,81, 023503

\bibitem[Hartlap \etal 2007]{hartlap}
Hartlap, J., Simon, P. \& Schneider, P. 2007, \aap \,464, 399

\bibitem[Hilbert \etal 2009]{hilbert}
Hilbert, S., Hartlap, J., White, S.~D.~M. \& Schneider, P. 2009, \,\aap 499, 31

\bibitem[Hu \& Sawicki 2007]{husawicki}
Hu, W. \& Sawicki, I. 2007, \prd \,76, 104043

\bibitem[Huterer \& Linder 2007]{huterer}
Huterer, D. \& Linder, E.~V. 2007, \prd \,75, 023519

\bibitem[Ilbert \etal 2009]{ilbert}
Ilbert, O., Capak, P., Salvato, M. \etal 2009, \apj \,690, 1236

\bibitem[Jain \& Khoury 2010]{jain}
Jain, B. \& Khoury, J. 2010, Annals of Phys. 325, 1479

\bibitem[Jain \& Zhang 2008]{jainz}
Jain, B. \& Zhang, P. 2008, \prd \,78, 063503 

\bibitem[Knox \etal 2006]{knox}
Knox, L., Song, Y.-S. \& Tyson, J.~A. 2006, \prd \,74, 023512

\bibitem[Komatsu \etal 2010]{komatsu}
Komatsu, E., Smith, K.~M., Dunkley, J. \etal 2010, arXiv:1001.4538

\bibitem[Koyama \etal 2009]{koyama}
Koyama, K. \& Taruya, A. \& Hiramatsu, T. 2009, \prd \,79, 123512

\bibitem[Lewis \& Bridle 2002]{lewis}
Lewis, A. \& Bridle, S. 2002, \prd \,66, 103511

\bibitem[Linder 2005]{linder}
Linder, E.~V. 2005, \prd \,72, 043529

\bibitem[Linder \& Cahn 2007]{lindercahn}
Linder, E.~V. \& Cahn, R.~N. 2007, Astropart. Phys. 28, 481

\bibitem[Lue \etal 2004]{lue}
Lue, A., Scoccimarro, R. \& Starkman, G.~D. 2004, \prd \,69, 124015

\bibitem[Ma \& Bertschinger 1995]{ma}
Ma, C.-P. \& Bertschinger, E. 1995, \apj \,455, 7

\bibitem[Massey \etal 2007]{massey07}
Massey, R., Rhodes, J., Leauthaud, A. \etal 2007, \apjs \,172, 239

\bibitem[Mukhanov \etal 1992]{mukhanov}
Mukhanov, V.~F., Feldman H.~A. \& Brandenberger, R.~H. 1992, Phys. Rep. 215, 206

\bibitem[Oyaizu \etal 2008]{oyaizu}
Oyaizu, H., Lima, M. \& Hu, W. 2008, \prd \,78 123524

\bibitem[Pogosian \etal 2010]{pogosian}
Pogosian, L., Silvestri, A., Koyama, K. \& Zhao, G.-B. 2010, \prd \,81, 104023

\bibitem[Schmidt 2008]{schmidt}
Schmidt, F. 2008, \prd \,78, 043002

\bibitem[Schrabback \etal 2010]{schrabback}
Schrabback, T., Hartlap, J., Joachimi, B. \etal 2010, \aap \,516, 63

\bibitem[Silvestri \& Trodden 2009]{silvestri}
Silvestri, A. \& Trodden, M. 2009, Rep. Prog. Phys. 72, 096901

\bibitem[Smith \etal 2003]{smith}
Smith, R.~E., Peacock, J.~A., Jenkins, A. \etal 2003, \mnras \,341, 1311

\bibitem[Song \& Dor{\'e} 2009]{song}
Song, Y.-S. \& Dor{\'e}, O. 2009, \jcap  \,3, 25

\bibitem[Song \etal 2007]{songfr}
Song, Y.-S., Hu, W. \& Sawicki, I. 2007, \prd \,75, 044004

\bibitem[Song \etal 2010]{songbd}
Song, Y.-S., Hollenstein, L., Caldera-Cabral, G. \& Koyama, K. 2010, \jcap \,4, 18

\bibitem[Thomas \etal 2009]{thomas}
Thomas, S.~A., Abdalla, F.~B. \& Weller, J. 2009, \mnras \,395, 197

\bibitem[Uzan 2007]{uzan}
Uzan, J.-P. 2007, Gen. Rel. Grav. 39, 307

\bibitem[Uzan \& Bernardeau 2001]{uzanb}
Uzan, J.-P., \& Bernardeau, F. 2001, \prd \,64, 083004i 

\bibitem[Wang \& Steinhardt]{wang}
Wang, L. \& Steinhardt, P.~J. 1998, \apj \,508, 483

\bibitem[Will 2006]{will}
Will, C. 2006, Living Rev. Rel. 9 - 3

\bibitem[Zhang \etal 2007]{zhang}
Zhang, P., Liguori, M., Bean, R. \& Dodelson, S. 2007, \prl \,99, 141302

\bibitem[Zhao \etal 2009]{zhaoa}
Zhao, G.-B., Pogosian, L., Silvestri, A. \& Zylberberg, J. 2009, \prd \,79, 083513

\bibitem[Zhao \etal 2010]{zhaotom}
Zhao, G.-B., Giannantonio, T., Pogosian, L. \etal 2010, \prd \,81, 103510

\end{thebibliography}
\end{document}